\documentclass[aps,prd,showpacs,amssymb,,twocolumn,nofootinbib]
{revtex4}
\usepackage{graphicx}
\begin{document}

\title{Quantum Time Uncertainty in a Gravity's Rainbow Formalism}

\author{Pablo \surname{Gal\'an}}
\affiliation{Instituto de Estructura de la Materia, C.S.I.C.,
Serrano 121, 28006 Madrid, Spain}

\author{Guillermo A. \surname{Mena Marug\'an}}
\affiliation{Instituto de Estructura de la Materia, C.S.I.C.,
Serrano 121, 28006 Madrid, Spain}

\begin{abstract}

The existence of a minimum time uncertainty is usually argued to
be a consequence of the combination of quantum mechanics and
general relativity. Most of the studies that point to this result
are nonetheless based on perturbative quantization approaches, in
which the effect of matter on the geometry is regarded as a
correction to a classical background. In this paper, we consider
rainbow spacetimes constructed from doubly special relativity by
using a modification of the proposals of Magueijo and Smolin. In
these models, gravitational effects are incorporated (at least to
a certain extent) in the definition of the energy-momentum of
particles without adhering to a perturbative treatment of the back
reaction. In this context, we derive and compare the expressions
of the time uncertainty in quantizations that use as evolution
parameter either the background or the rainbow time coordinates.
These two possibilities can be regarded as corresponding to
perturbative and non-perturbative quantization schemes,
respectively. We show that, while a non-vanishing time uncertainty
is generically unavoidable in a perturbative framework, an
infinite time resolution can in fact be achieved in a
non-perturbative quantization for the whole family of doubly
special relativity theories with unbounded physical energy.

\end{abstract}


\maketitle
\renewcommand{\theequation}{\arabic{equation}}

\section{Introduction}

In quantum mechanics, the passage of time can be tracked by
studying the evolution of the probability densities of observables
in a given quantum state \cite{GP}. Nevertheless, every observable
$\widehat{A}$ of the system has a characteristic time $\Delta_A t$
that limits the ability to detect its evolution, and that can be
estimated as the lapse needed by its expectation value $\langle
\widehat{A}\rangle$ to change an amount equal to its
root-mean-square (rms) deviation $\Delta A$, namely $\Delta_A
t\geq\Delta A/|d_t \langle \widehat{A}\rangle|$. On the other
hand, the quantum evolution of any explicitly time-independent
observable is given by Heisenberg equation $i\hbar\, d_t
\widehat{A}=[\widehat{A},\widehat{H}]$, where $\widehat{H}$ is the
Hamiltonian. Taking into account these expressions, together with
the uncertainty principle applied to the pair of observables
$\widehat{A}$ and $\widehat{H}$, and allowing the choice of any
observable $\widehat{A}$ of the system, one easily concludes that
any measurement of time made with our quantum state will have an
uncertainty $\Delta t$ (at least equal to the minimum of all
characteristic times $\Delta_A t$) that satisfies the inequality
$\Delta t\Delta H\geq \hbar/2$ \cite{GP}. This is usually called
the fourth Heisenberg relation.

Therefore, to improve the time sensitivity, states with a larger
and larger energy uncertainty must be allowed. However, in general
relativity, an uncertainty in the energy of the system implies an
uncertainty in the geometry. The latter introduces in turn an
uncertainty in the physical (or proper) time, if this corresponds
to a unit (asymptotic) timelike Killing vector of the metric
\cite{BMV,Luis}. In this way, the time uncertainty gets
contributions both from a purely quantum mechanical and from a
gravitational origin \cite{Luis}. As a consequence, an infinite
time resolution seems impossible, unless both types of
contributions are related in a very specific manner. Moreover,
since the energy of the system is generally defined in terms of
the (assumed) unit timelike Killing vector, the back reaction
leads also to a redefinition of the physical energy, thus giving
rise to new energy uncertainties. This non-trivial inter-twinning
between time and energy uncertainties in the presence of gravity
complicates the analysis of quantum measurements.

A way to face this problem is by adopting perturbative approaches,
in which one starts with a flat background and introduces in it
the matter content of the system, deforming hence the spacetime
geometry. This deformation subsequently results in a change of the
physical matter energy, leading to successive corrections in a
feed-back mechanism. Several arguments strongly support the idea
that this type of perturbative quantization always leads to a
minimum time uncertainty (at least in the next-to-leading order
approximation) \cite{Luis,Pad,unc}. However, it is not clear at
all whether a minimum time structure would emerge if one performed
the quantization of the gravitational system by adopting
non-perturbative schemes. This kind of schemes, for instance,
could allow one to encode in the theory, from the very beginning,
the modification of the physical energy-momentum of the matter
content owing to the process of back reaction.

In a recent paper \cite{BMV}, the quantum limits for time
resolution have been studied from both (perturbative and
non-perturbative) points of view in a family of gravitational
models that include the Einstein-Rosen (ER) cylindrical waves
\cite{ERwaves,ERscalar,ERash,ERBMV}. It has been shown that, in
these models, a minimum time uncertainty always exists if the
physical energy is bounded from above, as it happens to be the
case at least for ER waves \cite{ERash,ERbound}. Nonetheless, the
possibility was open that there could exist gravitational systems
with similar properties as those analyzed in that work but with an
unbounded physical energy. In these circumstances, it was argued
that an infinite time resolution could be reached in a
non-perturbative quantum description.

Moreover, for the systems considered in Ref. \cite{BMV}, the
behavior of the time uncertainty is radically different depending
on whether the quantization employs as evolution parameter either
a fixed time coordinate $T$ associated with a classical
(Minkowski) background or, alternatively, the physical time $t$,
which (for ER waves) coincides with the proper time in the
asymptotic region at spatial infinity. In the following, we will
understand by perturbative and non-perturbative quantizations
those quantum theories whose evolution is described, respectively,
in terms of these two types of time parameters, $T$ and $t$. The
motivation for this terminology is clear, since the time $T$ is
linked to a background solution, while $t$ is the physical time
whose definition includes the effects of the energy content on the
geometry. For the models considered in Ref. \cite{BMV}, the
relation between these two times is given by a scaling that
depends only on the energy of the solution (the energy of the
gravitational waves in the case of the ER spacetimes
\cite{ERash,ERBMV}).

It has also been proved recently \cite{2+1} that, from the
perspective of an equivalent formulation of the ER geometries as a
massless scalar field coupled to gravity in 2+1 dimensions
\cite{ERscalar}, these cylindrical waves can be viewed as an
example of the so-called doubly special relativity (DSR) theories
\cite{Amelino}. Such theories incorporate modifications to the
expressions of the energy and momentum of relativistic particles
owing to (possibly quantum) gravitational effects in such a way
that Lorentz symmetry is maintained but its implementation becomes
non-linear, so that it may be compatible with the presence of an
invariant scale in energy and/or momentum, ultimately related to
the Planck scale \cite{Amelino,DSR1,DSR2,DSR12}. Because of these
properties and the commented connection with ER waves, DSR
theories are natural candidates when trying to extend the
discussion presented in Ref. \cite{BMV} about the emergence of a
minimum time uncertainty in the presence of gravity.

In order to carry out this extension, an extra piece of
information must be added to the usual formulation of DSR theories
in momentum space, namely, the dual realization of these
relativity theories in position space. We will introduce a
modification of the gravity's rainbow proposal put forward by
Magueijo and Smolin \cite{rainbow}. This modification will ensure
the invariance of the symplectic structure defined in standard
special relativity, which can then be interpreted as corresponding
to a Minkowski background before switching on any gravitational
interaction. In this way, we will arrive at flat spacetime
coordinates that are related to those of the background by means
of a linear transformation which depends on the matter
energy-momentum. As a result, the metric associated with them can
be regarded as energy and momentum dependent. It is in this sense
that the so-constructed DSR theories can be considered a kind of
gravity's rainbow \cite{rainbow}.

We will show that, for this gravity's rainbow formalism, the
uncertainty in the physical time (conjugate to the physical
energy) is always strictly positive in perturbative quantization
schemes that employ as evolution parameter the time coordinate of
the auxiliary, flat background. However, an infinite time
resolution can actually be reached in a non-perturbative
quantization if the DSR theory involves an invariant momentum
scale, but not an energy scale. This example should clarify that
the emergence of a minimum time uncertainty in gravity is not
ineluctable in principle if one adopts a non-perturbative
quantization.

The rest of the paper is organized as follows. In Sect. II we
briefly review some results about DSR theories, formulated in
momentum space. We describe the relation between the physical
energy-momentum and the pseudo energy-momentum, on which the
Lorentz transformations act linearly. This relation is provided by
a non-linear map $U$ whose properties we discuss. Sect. III deals
with the dual realization of the DSR theories in position space.
We derive the expressions for the spacetime coordinates that are
conjugate to the physical energy-momentum. Assuming an underlying
Hamiltonian framework, we then analyze the quantization of this
gravity's rainbow formalism. In Sect. IV we obtain the uncertainty
in the physical time for a perturbative quantization, proving that
it cannot vanish under very mild hypotheses. In Sect. V we
demonstrate that, on the contrary, the uncertainty in the physical
time can be as small as desired in a non-perturbative
quantization, provided that the DSR theory has no invariant energy
scale corresponding to a maximum of the physical energy. Finally,
Sec. VI contains the conclusions and some further discussion. In
the following, all dimensionful quantities will be expressed in
Planck units. In particular, we set $\hbar=c=1$.

\section{DSR in momentum space}

DSR theories are characterized by a non-linear action of the
Lorentz transformations in momentum space that preserves an energy
or momentum scale (besides respecting the role of the speed of
light as a fundamental scale) \cite{Amelino,DSR1,DSR2,DSR12}. A
way to understand this non-linear action is by mapping the
physical energy-momentum $P^{a} = (E,p^{i})$ into a standard
Lorentz 4-vector $\Pi^{a} = (\epsilon,\pi^{i})$, which transforms
in a linear way \cite{judes}. The involved non-linear map is
generally denoted by $U$, and the 4-vector $\Pi^a$ is called the
pseudo energy-momentum. Lowercase Latin indices from the beginning
and the middle of the alphabet denote, respectively, Lorentz and
(flat) spatial indices. The map $U$ must be invertible; then the
transformation of the physical energy-momentum is given by
\cite{judes,MS}
\begin{equation}\label{nonlin}L(P) = (U^{-1}\circ\mathcal{L}\circ U)
(P),\end{equation} where $\mathcal{L}$ is the standard linear
action of the Lorentz transformation.

In the sector of small energies and momenta compared to the DSR
scale, the physical and pseudo variables must coincide and,
therefore, the map $U$ must reduce to the identity, a property
that will be used in the following. In addition, it is usually
assumed that the standard action of rotations is not modified in
DSR theories \cite{MS,kappa}. As a consequence, the most general
functional form of $U$ (and of its inverse) is \cite{kappa}
\begin{eqnarray}\label{momenta}\Pi = U(P) &\Rightarrow&
\left\{\begin{array}{l} \epsilon = \tilde{g}(E,p)\,,\\ \pi^{i} =
\tilde{f}(E,p)\,\frac{p^{i}}{p}\,,
                       \end{array}
               \right. \nonumber\\
P = U^{-1}(\Pi) &\Rightarrow& \left\{\begin{array}{l} E =
g(\epsilon,\pi)\,,\\ p^{i} =
f(\epsilon,\pi)\,\frac{\pi^{i}}{\pi}\,,
                       \end{array}\right.\end{eqnarray}
where $p:=|\vec{p}\,|$ and $\pi:=|\vec{\pi}|$. So the map $U$ is
totally determined by two scalar functions $\tilde{g}$ and
$\tilde{f}$ (or $g$ and $f$).

Since standard Lorentz boosts run over the whole range
$[0,\infty)$ for both energy and (the norm of the) momentum, the
image of $U$ must equal this range, so that the inverse of
$\mathcal{L}\circ U$ can always exist in Eq. (\ref{nonlin}).
Furthermore, in order to have a finite energy scale $E^{*}$
(and/or momentum $p^{*}$) invariant under the Lorentz
transformations (\ref{nonlin}), it is necessary that the map $U$
sends it to infinity in the space of pseudo energy-momentum
vectors, since this is the only invariant scale in standard
special relativity. Therefore, the map $U$ must be singular at
$E^{*}$ (and/or $p^{*})$ and the domain of definition of $U$
(assumed to contain the sector of low energies) is bounded by that
scale \cite{MS}. We then have three possible types of DSR
theories, depending on whether one has only a bounded physical
momentum (DSR1 type), a bounded physical energy (DSR3 type), or
bounds in both physical quantities (DSR2 type).

More explicitly, if we consider a particle with pseudo mass
$\mu\geq 0$ (namely, the Casimir invariant of the pseudo momentum
space $\mu^2=\epsilon^2-\pi^2$, related to the rest mass $m_0$ by
$\mu=\tilde{g}(m_0,0)$ \cite{judes}), then, in the limit of
infinite momentum on the mass shell (denoted by $
\pi|_{\mu}\rightarrow\infty$), the existence of an invariant
scale, where the map $U$ is singular, implies one (or both) of the
following possibilities;
\begin{eqnarray}\label{bounds}
                {\rm a)}&&\;\; \lim_{\pi|_{\mu}\rightarrow\infty}
                g=E^{*}<\infty\,,\nonumber\\
                {\rm b)}&&\;\; \lim_{\pi|_{\mu}\rightarrow\infty}
                f=p^{*}<\infty.
\end{eqnarray}
Possibility a) is realized for DSR2 and DSR3 types of theories,
but not for DSR1. On the other hand, the behavior b) is found only
in the DSR1 and DSR2 classes. In general, the invariant scale is
assumed to be of the Planck order, but this supposition, motivated
by quantum considerations, can be relaxed.

\section{A gravity's rainbow proposal}

The recent interest in deformed dispersion relations, justified by
their potential observational consequences in fields like
astrophysics \cite{phenomen}, explains why DSR theories are
usually formulated in momentum space. Within this formulation, the
transformation laws in position space are not determined. There
exist different proposals for constructing a modified spacetime
geometry consistent with DSR \cite{kappa,position}. One of them,
suggested by several hypotheses concerning quantum gravity,
consists in introducing a non-commutative geometry, namely,
admitting that spacetime coordinates no longer commute
\cite{DSR12,kappa}. An example of this is the $\kappa$-deformed
Minkowski spacetime. However, non-commuting spacetime coordinates
are not a necessary consequence of DSR theories: the realization
in position space can be achieved in the framework of commutative
geometries \cite{kappa,position,mignemi}.

For instance, a way to specify this realization was recently
proposed by Magueijo and Smolin \cite{rainbow}. By demanding that
the contraction between the energy-momentum and an infinitesimal
spacetime displacement be a linear invariant, they derived
modified expressions for the spacetime coordinates that are linear
in the original (Minkowski) background coordinates $q^a$, but
depend non-trivially on the energy-momentum. Owing to this
dependence, a rainbow of metrics emerged in the formalism, each
particle being associated with a different metric according to its
energy-momentum.

Here, we will adopt a related kind of proposal, but, instead of
the above contraction, we will demand the invariance of the
symplectic form ${\bf d}q^a\wedge {\bf d}\Pi_a$ [where
$\Pi_a=(-\epsilon,\pi^i)$ and the wedge denotes the exterior
product for differential forms]. The modified position variables
$x^a$ obtained in this way are then conjugate to the physical
energy-momentum $P_a$, i.e., the map from $(q^a,\Pi_a)$ to
$(x^a,P_a)$ is just a canonical transformation. The physical
energy-momentum can then be assigned the role of generator of
spacetime translations in the coordinates $x^a$. In fact, the same
requirement of covariance, ensuring that the space of coordinates
can be identified with the cotangent space for the physical
energy-momentum, was already put forward by Mignemi \cite{mignemi}
(though introduced in a different manner).

An additional reason supporting the suggested change with respect
to Ref. \cite{rainbow} is that it leads to the correct expression
for the physical time (and spatial coordinates) in the case of ER
waves (formulated in 2+1 dimensions) \cite{ERBMV,BMV}, as we will
in part discuss later. Since this and other physical implications
of our proposal significantly differ from those of the formalism
presented in Ref. \cite{rainbow}, one can view our construction as
a distinct realization of DSR theories in position space, rather
than simply as a modification. Nevertheless, it is worth
commenting that the essential feature employed in the rest of our
analysis is that the relation between the background and the
physical (rainbow) spacetime coordinates is a linear
transformation that depends only on the energy-momentum. This
property persists even if one adheres exactly to the Magueijo and
Smolin proposal, the only difference being the detailed form of
the transformation.

It is straightforward to complete the map $U$ in momentum space
into a contact canonical transformation providing position
variables conjugate to $P_a$. Employing the form of this map from
$P_a=(-E,p_i)$ to \begin{equation}\Pi_a =\left(-\tilde{g}(E,p),\,
\tilde{f}(E,p)\frac{p_i}{p}\right),\end{equation} it is easy to
see that the desired transformation is generated by the function
\begin{equation}\label{generate} F(q^a,P_b)=-\tilde{g}(E,p)q^{0}\,
+\,\tilde{f}(E,p)\frac{p_jq^{j}}{p}.
\end{equation}
Then, $x^a=\partial F/\partial P_a$. Making use of the implicit
function theorem (and the identity $p_j/p=\pi_j/\pi$), we finally
get the expressions for the new spacetime coordinates:
\begin{eqnarray}\label{x}x^{0}&=&\frac{1}{{\rm det}
J(\epsilon,\pi)}\left[\frac{\partial
f(\epsilon,\pi)}{\partial\pi}q^{0}+\frac{\partial
f(\epsilon,\pi)}{\partial \epsilon}\frac{\pi_i}{\pi}q^{i}\right],
\nonumber\\ x^{i}&=&\frac{1}{{\rm
det}J(\epsilon,\pi)}\left[\frac{\partial g(\epsilon,\pi)}{\partial
\pi}\frac{\pi^{i}}{\pi}q^{0}+\frac{\partial
g(\epsilon,\pi)}{\partial\epsilon}\frac{\pi^{i}
\pi_j}{\pi^2}q^{j}\right] \nonumber\\ &&+\frac{\pi}
{f(\epsilon,\pi)}\left(q^i-\frac{\pi^{i}\pi_j}{\pi^2}q^{j}\right).
\end{eqnarray}
Here, $g$ and $f$ are the two functions that fix the inverse map
$U^{-1}$, and \begin{equation}{\rm det}J=\frac{\partial
g}{\partial\epsilon}\frac{\partial f} {\partial\pi}-\frac{\partial
g}{\partial\pi}\frac{\partial f}{\partial\epsilon}.\end{equation}

In the following, we will call physical variables to the canonical
set formed by $x^a$ and the physical energy-momentum, whereas we
will refer to $q^a$ and $\pi_a$ as background or auxiliary
variables. In addition, to simplify in part our index notation, we
will designate $q^0$ by $T$ and $x^0$ by $t$ (this type of
notation reproduces that employed in Ref. \cite{BMV}). Finally we
note that, as we commented that happens for the energy-momentum,
the physical and background coordinates coincide in the limit
where energies and momenta are small compared to the DSR scale,
since in this regime $g(\epsilon,\pi)\approx\epsilon$ and
$f(\epsilon,\pi)\approx\pi$.

\section{Physical time uncertainty: Perturbative case}

Let us assume that our system possesses an underlying Hamiltonian
formalism such that the values of the physical and pseudo energies
are determined, respectively, by a physical Hamiltonian $H$ and a
background one $H_0$. In agreement with our previous discussion,
in this Hamiltonian system the physical and pseudo momenta $p_i$
and $\pi_i$ are conjugate to the position variables $x^i$ and
$q^i$, whose translations they generate. In addition, motivated in
part by the fact that DSR theories are supposed to provide
effective descriptions of free particles, we also assume that our
system is free, so that the energy and momentum are conserved (had
one to consider composite systems, the physical energy and
momentum would not be additive). In this way, apart from being
time independent, the Hamiltonian must indeed commute under
Poisson brackets with the momentum, both for the physical and the
background variables.

From Eq. (\ref{momenta}), we have that $E \rightarrow
H=g(H_0,\pi)$ and $\epsilon\rightarrow H_0=\tilde{g}(H,p)$. In
this section, we will analyze the quantization of the system with
evolution generated by the background Hamiltonian $H_0$. In such a
quantization, the evolution parameter is the corresponding time
coordinate $q^0=T$, namely, the background time. We leave for Sec.
V the analysis of the alternative quantization with evolution
parameter given by $x^0=t$.

\subsection{Calculation of the time uncertainty}

Let us admit that a quantization of the system with evolution
generated by the background Hamiltonian $H_0$ is feasible. In this
perturbative quantization, the background time $T$ plays the role
of evolution parameter, whereas the physical time is in fact
promoted to an operator $\widehat{t}$ \cite{BMV}. Taking into
account the expression of $x^{0}=t$ obtained in Eq. (\ref{x}), and
replacing energies by Hamiltonians, we can write
\begin{eqnarray}\label{t}\widehat{t}&=&\widehat{A}(H_0,\pi)T+
\widehat{C}_T,
\\\label{C} \widehat{C}_T&=& \frac{\widehat{B}(H_0,\pi)
\widehat{Q}_T+\widehat{Q}_T\widehat{B}(H_0,\pi)}{2},\end{eqnarray}
where
\begin{eqnarray}
\label{A} A(H_0,\pi)&=&\frac{1}{{\rm det} J(H_0,\pi)}
\frac{\partial f(H_0,\pi)}{\partial\pi},\\ \label{B} B(H_0,\pi)&=&
\frac{1}{{\rm det} J(H_0,\pi)}\frac{\partial f(H_0,\pi)}{\partial
H_0},\\ \label{Q}Q_T&=&\frac{\pi_i q^i}{\pi}.\end{eqnarray} In Eq.
(\ref{C}) we have symmetrized the product of $\widehat{B}$ and
$\widehat{Q}_T$ (although our results are insensitive to the
actual choice of factor ordering for this product) and the
operators $\widehat{A}$ and $\widehat{B}$ can be defined, using
the spectral theorem, in terms of those for the background
Hamiltonian and momentum ($H_0$ and $\pi$) which, according to our
comments above, are assumed to commute (so that the momentum is
conserved quantum mechanically). As for the operator representing
$Q_T$, we will analyze its form in brief. Let us simply remark for
the moment that it will generically be time dependent since, under
quantization, the auxiliary spatial variables $q^i$ will not
commute with the Hamiltonian. Therefore the sub-index notation for
the operators $\widehat{Q}_T$ and $\widehat{C}_T$. Note that, by
contrast, our assumptions guarantee that $\widehat{A}$ and
$\widehat{B}$ are time independent.

Given a quantum state, we can measure the probability densities of
the operators $\widehat{A}$ and $\widehat{C}_T$ \cite{nota0}. Let
us call $\Delta A$ and $\Delta C_T$ their rms deviations. In order
to evaluate the operator $\widehat{t}$, we still need to determine
the value of the parameter $T$. The passage of this time parameter
can be tracked by analyzing the evolution of probability densities
of observables in the quantum state. This process leads to a
statistical measurement of $T$, with probability density $\rho
(T)$. We denote the associated mean value by $\overline{T}$.
Obviously, the corresponding uncertainty in $T$ must satisfy the
fourth Heisenberg relation $\Delta T\Delta H_0\geq 1/2.$ With this
measurement procedure, the physical time uncertainty would be
\begin{eqnarray}\label{uncert}(\Delta t)^{2}&=&\int dT \rho (T)
\big\langle\left(\widehat{A}T+\widehat{C}_T-\langle\widehat{A}
\rangle\overline{T}-
\langle\widehat{C}_{\overline{T}}\rangle\right)^{2}
\big\rangle\nonumber\\ &=& \int dT \rho(T)\left\{T^{2}(\Delta
A)^{2} + \langle
\widehat{A}\rangle^2(T^2-\overline{T}^{2})\right.\nonumber\\
&&\hspace*{.8cm}+\left.T\langle\widehat{A}\widehat{C}_T+
\widehat{C}_T\widehat{A}\rangle-2 \overline{T}\langle
\widehat{A}\rangle\langle\widehat{C}_{T}\rangle\right.
\nonumber\\&&\hspace*{.8cm}+\left.\langle
\widehat{C}_T^{2}\rangle+\langle\widehat{C}_{\overline{T}}
\rangle^2-2\langle\widehat{C}_T\rangle\langle
\widehat{C}_{\overline{T}}\rangle\right\}.
\end{eqnarray}
Here, $\langle\widehat{O}\rangle$ denotes the expectation value in
our quantum state of any operator $\widehat{O}$. In addition, in
the estimation of the mean value of the physical time, we have
substituted the parameter $T$ by its corresponding mean value
$\overline{T}$ (in particular, $\widehat{C}_{\overline{T}}$ is the
operator $\widehat{C}_{T}$ at the instant $\overline{T}$)
\cite{nota}.

This expression becomes relatively simple when the dependence of
$\widehat{C}_T$ on $T$ is linear. In fact, this is the case with
our hypothesis that the system is free. To be more specific let us
accept, according to our hypothesis, that the Hamiltonian $H_0$ is
a scalar function of the pseudo momentum $\pi$ (and some
parameters). The assumed canonical symplectic structure for the
background variables implies that $Q_T$ [given by Eq. (\ref{Q})]
and $\pi$ are canonically conjugate, i.e., their Poisson bracket
is $\{Q_T,\pi\}=1$. Since $H_0$ generates the evolution in $T$,
one then has that, classically, $dQ_T/dT=\{Q_T,H_0\}=dH_0/d\pi$.
Obviously $dQ_T/dT$ is constant (because $dH_0/d\pi$ depends only
on the pseudo momentum, which is a conserved quantity), and
therefore $Q_T=Q_0+T\;(d H_0/d\pi)$. We can then promote $Q_T$ to
a linearly $T$-dependent observable by representing $Q_0$ as a
time independent operator and defining $d H_0/d\pi$ in terms of
the pseudo momentum operator by means of the spectral theorem.
Taking into account that $\widehat{B}(H_0,\pi)$ is constant in
time, Eq. (\ref{C}) shows then that $\widehat{C}_T$ is linear in
$T$.

The above analysis allows us to write the operator $\widehat{t}$
in the alternative form
\begin{eqnarray}\label{t2}\widehat{t}\!&=&\!
\widehat{V}(H_0,\pi)T+\widehat{W}(H_0,\pi,Q_0),
\\\label{V} \widehat{V}(H_0,\pi)\!&=&\!
\widehat{A}(H_0,\pi)+\widehat{B}(H_0,\pi)\widehat{\frac{d
H_0}{d\pi}}(\pi) ,\\ \label{W}\hspace*{-.5cm}
\widehat{W}(H_0,\pi,Q_0)\!&=&\!\frac{\widehat{B}(H_0,\pi
)\widehat{Q}_0+\widehat{Q}_0\widehat{B}(H_0,\pi)}{2}
.\end{eqnarray} In Eq. (\ref{V}) we have employed that $H_0$ and
$\pi$ commute as operators. We emphasize that, since
$\widehat{Q}_0$ is time independent, so are $\widehat{V}$ and
$\widehat{W}$.

Formula (\ref{uncert}) for the time uncertainty in the physical
time still applies, but now with $\widehat{A}$ identified with
$\widehat{V}$, and $\widehat{C}_T$ and
$\widehat{C}_{\overline{T}}$ substituted by $\widehat{W}$. The
result can be expressed in the form
\begin{equation}(\Delta t)^{2}=[\Delta(V\overline{T}+W)]^{2}
+\langle \widehat{V}\rangle^{2}(\Delta T)^{2}+(\Delta T\Delta
V)^{2}. \label{unc}\end{equation} Since we have the sum of three
positive terms in this equation, for the physical time uncertainty
to vanish it is necessary that all of them be zero.

Let us show that this will not generically happen. From the first
term in Eq. (\ref{unc}), one can easily see that the uncertainty
in the physical time vanishes for $T\gg 1$ if and only if $\Delta
V$ becomes equal to zero at large values of $T$. Since the
operator $\widehat{V}$ is time independent, its rms deviation
vanishes then at any instant of time $T$. Assume now that the
expression of the Hamiltonian in terms of $\pi$ is invertible in
the whole range of auxiliary energies, i.e. $\pi=\pi(H_0)$
\cite{nota2}, and define ${\cal V}[H_0]:=V[H_0,\pi(H_0)]$. An
alternative possibility is that $V$ is independent of $\pi$, in
which case we straightforwardly identify ${\cal V}$ with $V$. In
any of these cases, assume finally that $d{\cal V}/dH_0\neq 0$ for
all the allowed values of $H_0$, so that the correspondence
between $H_0$ and its image under ${\cal V}$ is one-to-one (a
similar assumption was made in Ref. \cite{BMV}). Making use of the
spectral theorem, the requirement that $\Delta V=\Delta {\cal V}$
vanish implies then that $\Delta H_0=0$, because our assumption
guarantees that the eigenstates of these two operators coincide.
In these circumstances, the fourth Heisenberg relation states that
$\Delta T$ is unbounded.

We will now show that the product of uncertainties $\Delta T\Delta
V=\Delta T \Delta {\cal V}$ that appears in Eq. (\ref{unc}) cannot
vanish when $\Delta H_0$ approaches zero, thus concluding the
proof that $\Delta t$ is strictly positive. Expanding ${\cal
V}(H_0)$ around the expectation value of $H_0$, where it is peaked
when $\Delta H_0$ is small, we arrive at
\begin{equation}(\Delta {\cal V})^2=\langle \widehat{\cal V}^{2}
-\langle \widehat{\cal V}\rangle^{2}\rangle \approx
\left(\left.\frac{d{\cal V}}{dH_0}\right|_{\langle
\widehat{H}_0\rangle}\Delta H_0\right)^2.\end{equation} Hence, in
the limit of localized energy,
\begin{equation}
\lim_{\Delta H_0\rightarrow 0}\!\Delta T\Delta {\cal V}\!\geq
\!\lim_{\Delta H_0\rightarrow 0} \frac{\Delta{\cal V}}{2\Delta
H_0}\!=\!\left|\frac{1}{2}\!\left.\frac{d{\cal
V}}{dH_0}\right|_{\langle \widehat{H}_0\rangle}\right|\!\neq
0.\end{equation} In conclusion, at least under very mild
assumptions, the uncertainty in the physical time cannot be zero
for an observer that describes the quantum evolution using as time
parameter the background time $T$.

As a particular example we can analyze the case of ER waves, where
the physical and pseudo momenta coincide, and the physical energy
is $E=(1-e^{-4\epsilon})/4$ (for an effective gravitational
constant in three-dimensions equal to the unity in Planck units)
\cite{BMV}. Introducing Hamiltonians, we thus have
$g(H_0,\pi)=(1-e^{-4H_0})/4$ and $f(H_0,\pi)=\pi$. Since $f$ is
energy independent, Eq. (\ref{B}) leads to $B=0$, a fact that
considerably simplifies the expressions of the physical time and
its uncertainty. From Eqs. (\ref{A}), (\ref{V}), and (\ref{W}), we
get $A=1/(dg/dH_0)=e^{4H_0}$, $\widehat{V}=e^{4\widehat{H_0}}$,
and $\widehat{W}=0$. Given that the deduced function $A$ (and
hence ${\cal V}$) is strictly increasing in $H_0$, the assumptions
introduced above are satisfied, and the conclusion of a non-zero
uncertainty in the physical time holds. In this way, one recovers
the results obtained in Ref. \cite{BMV}.

\subsection{First order corrections}

In this subsection, we will analyze the behavior of the
uncertainty in the physical time when one approximates this
operator by keeping only up to first order corrections in the
energy. We will see that the results lend additional support to
the statement that this uncertainty is strictly positive in the
perturbative approach to the quantization.

In order to study the desired corrections, we start by expanding
the functions $g(H_0,\pi)$ and $f(H_0,\pi)$ around the minimum of
the pseudo energy and around vanishing pseudo momentum. We will
denote the minimum pseudo energy by $\mu$, motivated by the
standard relativity case, where it equals (the square root of) the
Casimir invariant, $\mu^2=\epsilon^2-\pi^2$. We assume that the
functions $g$ and $f$ are smooth and that $\mu$ is small compared
to the invariant DSR scale(s). In particular, this last fact
allows us to employ that, to leading order, $g(H_0,\pi)\approx
H_0$ and $f(H_0,\pi)\approx \pi$ in the region of the expansion.
To derive the first order corrections to the physical time, in the
expansion of $g$ and $f$ it is actually necessary to keep only up
to quadratic terms in the variables $\pi$ and
\begin{equation}{\cal H}_0:=H_0-\mu.\end{equation} One can then
use Eqs. (\ref{A}) and (\ref{B}) to obtain the expressions of
$A(H_0,\pi)$ and $B(H_0,\pi)$ up to linear terms in those
variables:
\begin{eqnarray}\label{ABFO}
A&\approx& 1-\left.\frac{\partial^{2}g}{\partial H_0^{2}}
\right|_0 {\cal H}_0-\left.\frac{\partial^{2}g}{\partial
H_0\partial\pi}\right|_0 \pi , \nonumber\\
B&\approx&\left.\frac{\partial^{2}f}{\partial H_0^{2}} \right|_0
{\cal H}_0+\left.\frac{\partial^{2}f}{\partial H_0
\partial\pi}\right|_0 \pi ,\end{eqnarray}
where the symbol $\,\!|_0$ stands for evaluation at $H_0=\mu$ and
$\pi=0$.

Next, from Eqs. (\ref{V}) and (\ref{W}) one can easily calculate
the first order corrections to the leading behavior of
$\widehat{V}$ and $\widehat{W}$. In this step, one needs to
introduce the expression of the Hamiltonian in terms of the
momentum, $H_0(\pi)$ [see Eq. (\ref{V})]. On the one hand, it is
natural to assume that the minimum of the pseudo energy is reached
for vanishing pseudo momentum, $H_0(0)=\mu$. On the other hand,
motivated by the standard relativity case
[$\epsilon=\sqrt{\mu^2+\pi^2}\rightarrow H_0(\pi)$], two cases are
worth considering.

1) {\it ``Massive'' case}: $\mu\neq0$, with
$(dH_0/d\pi)|_{\pi=0}=0$.\vspace{.1cm}

We get $H_0(\pi)\approx\mu+b \pi^2$, where
$2b=(d^{2}H_0/d\pi^{2})|_{\pi=0}.$ Assuming that $b> 0$, it
follows that $\pi\approx \sqrt{{\cal H}_0/b}$. For instance, in
standard special relativity one would have $b=1/(2\mu)$.
Corrections linear in ${\cal H}_0$ are hence negligible compared
to those proportional to $\pi$. In addition, $dH_0/d\pi\approx
2b\pi$, which can be neglected compared to the unity. As a
consequence, we arrive at the following approximations at
next-to-leading order:
\begin{eqnarray}\label{massV}
\widehat{V}&\approx & \widehat{A}\approx 1-
\left.\frac{\partial^{2}g}{\partial H_0
\partial\pi}\right|_0\frac{\widehat{\cal H}_0^{1/2}}{\sqrt{b}}
,\\ \label{massW} \widehat{W}&\approx &
\left.\frac{\partial^{2}f}{\partial H_0
\partial\pi}\right|_0\frac{\widehat{\cal H}_0^{1/2}
\widehat{Q}_0+\widehat{Q}_0 \widehat{\cal H}_0^{1/2}}
{2\sqrt{b}}.\end{eqnarray}
The physical time uncertainty in this
approximation can be obtained from Eq. (\ref{unc}).

Note that the resulting leading term (zeroth order in the energy)
is the uncertainty of the background time in standard quantum
mechanics. We also point out that the function ${\cal V}$,
introduced in the previous subsection, is given in the studied
approximation just by (the classical counterpart of) Eq.
(\ref{massV}). Such a function is clearly monotonic in $H_0$ (or
${\cal H}_0$), provided that the second partial derivative
$\left.[\partial^{2}g/(\partial H_0
\partial\pi)]\right|_0$ does not vanish, so that we have
really kept the first order energy corrections to ${\cal V}$.
Then, the assumptions made at the end of Subsec. IV.A hold,
leading to the conclusion that the physical time uncertainty
cannot be made zero.

2) {\it ``Massless'' case}: $\mu\!=\!0$, with
$(dH_0/d\pi)|_{\pi=0}\!=k\neq 0$.\vspace*{.1cm}

\noindent In this case $H_0\approx k\pi$ and ${\cal H}_0=H_0$. In
standard special relativity, for instance, one would have $k=1$.
Corrections linear in $H_0$ and in $\pi$ are then of the same
order, and $dH_0/d\pi$ is of order unity. Therefore, one obtains
in the linear order approximation:
\begin{eqnarray}
\widehat{V}&\approx & 1+
\left(-\left.\frac{\partial^{2}g}{\partial H_0^2}\right|_0-
\frac{1}{k}\left.\frac{\partial^{2}g}{\partial H_0
\partial\pi}\right|_0+k\left.\frac{\partial^{2}f}
{\partial H_0^2}\right|_0\right.\nonumber\\
&&\hspace*{.9cm}+\left.\left. \frac{\partial^{2}f}{\partial
H_0\partial\pi}\right|_0\right) \widehat{H}_0 , \label{mlV}\\
\label{mlW} \widehat{W}&\approx &
\left(\left.\frac{\partial^{2}f}{\partial H_0^2}\right|_0\!+
\frac{1}{k}\left.\frac{\partial^{2}f}{\partial H_0
\partial\pi}\right|_0\right)\frac{\widehat{H}_0
\widehat{Q}_0+\widehat{Q}_0\widehat{H}_0}{2}.
\end{eqnarray}
At this order, the function ${\cal V}$ is approximated by (the
classical analog of) Eq. (\ref{mlV}). Accepting that the
coefficient in front of $H_0$ in that expression is non-zero, so
that we have actually included the next-to-leading order
correction, we arrive again to a monotonic function of $H_0$.
Hence, the line of reasoning discussed in Subsec. IV.A applies,
and we conclude that it is impossible to reach the limit of
infinite resolution in the physical time.

\section{Physical time uncertainty: Non-perturbative case}

We turn now to the discussion of the physical time uncertainty
when one adopts the point of view that the quantum evolution of
the system is generated by the physical Hamiltonian $H$. It is
worth commenting that, if the system admits a perturbative
quantization where the background Hamiltonian $H_0$ and momentum
$\pi$ are promoted to self-adjoint operators, a non-perturbative
quantization is also possible. To see this, notice that, in the
representation employed for the perturbative quantization, the
spectral theorem allows one to define as self-adjoint operators
the physical Hamiltonian $H$ and momentum $p$, given by the
functions $g$ and $f$ in terms of $H_0$ and $\pi$. The
exponentiation of this operator realization of $H$ provides then a
unitary evolution operator, that describes the dynamics in a time
parameter that can be identified with the physical time $t$.
Clearly, in the so-constructed non-perturbative quantization, the
uncertainty of $t$ is only limited by the fourth Heisenberg
relation, taking as Hamiltonian the physical one, namely $\Delta
t\Delta H\geq 1/2$.

As a consequence, for an observer in the non-pertur\-bative
quantum system, the resolution for the physical time is
intrinsically bounded if and only if the same happens for the
physical energy (i.e., the physical Hamiltonian). The conclusion
does not depend on other details of the system. The only relevant
point is whether the range of the physical energy is infinite.
This range is determined by the image of $g$, one of the two
functions that characterize the DSR theory. But the image of $g$
is bounded from above if and only if the DSR theory possesses an
invariant energy scale [remember Eq. (\ref{bounds})]. This is not
always the case: it occurs only in the so-called DSR2 and DSR3
types of theories, but not for the DSR1 class. Therefore, a finite
time resolution is not a necessary consequence of the quantization
of the system, at least in this non-perturbative framework. More
specifically, for the whole family of DSR1 theories
\cite{Amelino,DSR1,kappa}, where only an invariant scale in
momentum exists, the quantum resolution in the physical time can
be made (non-pertubatively) as large as desired.

\section{Summary and discussion}

We have investigated the existence of a minimum time uncertainty
in a modified gravity's rainbow formalism, obtained by means of a
dual realization of DSR theories in position space. This
realization leads to a set of spacetime coordinates that are
canonically conjugate to the physical energy-momentum. Such
coordinates are constructed from the (Minkowski) background ones
by means of a linear transformation that depends on the energy and
momentum. Assuming an underlying Hamiltonian formulation, with
energy determined by the value of the generator of the evolution,
and concentrating our attention on free systems, we have discussed
the differences in adopting as dynamical generator either the
physical or the background Hamiltonian, the latter corresponding
to the pseudo energy.

If the dynamics is dictated by this last Hamiltonian, the
evolution parameter of the quantum theory is the background time
$T$, and the physical time $t$ is described by a $T$-dependent
family of operators. We have shown that its uncertainty cannot be
made to vanish, at least under very mild assumptions about the
features of the background Hamiltonian and the DSR theory. In
fact, these assumptions are only sufficient, but not necessary in
order to prove that the studied uncertainty is greater than zero.
For instance, one can show that the resolution in the physical
time is finite as well for all those cases in which the function
${\cal V}$ is strictly positive (so that the background and
physical arrows of time coincide) and the ratio ${\cal
V}(H_0)/H_0$ has a non-zero limit when $H_0$ tends to infinity (so
that, in the high energy sector, ${\cal V}$ grows at least like
$H_0$ by a constant). Therefore, an infinite resolution in the
physical time cannot (generically) be reached within a
quantization framework in which the energy-momentum modifications
in the definition of time are not incorporated in the choice of
evolution parameter.

By contrast, when the quantum dynamics is generated by the
physical energy, the role of evolution parameter is directly
assigned to the physical time. In this case, its uncertainty is
only limited by quantum mechanics via the fourth Heisenberg
relation. As a result, an infinite resolution is possible if and
only if the physical energy of the system is unbounded from above,
which in turn is equivalent to the absence of an invariant energy
scale in the DSR theory. There exists a whole family of DSR
theories that possess a momentum scale but not an energy scale of
this kind, namely, the so-called DSR1 theories, whose prototype is
a model suggested by Amelino-Camelia \cite{Amelino,DSR1}. This
clearly demonstrates that, in non-perturbative quantum
descriptions, the existence of a minimum uncertainty in the
physical time is not generally unavoidable when gravitational
effects are taken into account.

An issue for further discussion is whether, in those
non-perturbative quantum systems where an infinite time resolution
is possible, there emerges, nonetheless, a minimum uncertainty in
the spatial position, as could be suggested by the presence of a
bound for the physical momentum in DSR1 theories, supplied by the
invariant scale. We plan to study this question in the future, as
a natural continuation of the analysis carried out here.

Our discussion can be regarded as a generalization of that of Ref.
\cite{BMV}. Apart from the hypotheses concerning the existence of
a feasible quantization and the recovery of the standard results
in the low energy sector, the rest of conditions assumed for the
models studied in Ref. \cite{BMV} amount to accept a relation
between physical and background coordinates of the form (\ref{x}),
but with the DSR functions $f$ and $g$ satisfying: i) $f$ is
independent of the pseudo energy, and ii) $g$ is a convex or
concave (invertible) smooth function of only the pseudo energy. In
these cases, one can check that, with our notation,
\begin{equation}{\cal
V}(H_0)=V(H_0)=A(H_0)=\frac{1}{dg/dH_0}\end{equation}
and\begin{equation} \frac{d{\cal
V}}{dH_0}=\frac{dV}{dH_0}=-\frac{d^2g/dH_0^2}{(dg/dH_0)^2}\neq 0.
\end{equation}
Therefore, the assumptions introduced at the end of Subsec. IV.A
hold in these models, and thus $\Delta t$ cannot be made equal to
zero in the perturbative quantization.

Finally, in our analysis we have implicitly kept in mind the case
of a relativistic particle, motivated by the formulation of DSR
theories as alternatives to special relativity (at least in
momentum space). Since a field can be viewed as a combination of
particles, one might try to extend the arguments presented here to
a quantum field theory context. In perturbative quantum field
theories, the background space coordinates $q^{j}$ should be
treated as parameters. Therefore, one would expect that the
physical time operator adopted an expression of the form
$\widehat{t}=\widehat{A}T+\widehat{D}_jq^{j}$ [see Eq. (\ref{t})].
Then, the resulting time uncertainty would be
\begin{eqnarray}(\Delta t)^{2}&=&
\left[\Delta\left(\sum_a D_a\overline{q^{a}}\right)\right]^{2}
+\sum_a\langle\widehat{D}_a\rangle^2(\Delta
q^{a})^{2}\nonumber\\&&+\sum_a(\Delta q^{a})^2 (\Delta D_a)^{2},
\end{eqnarray}
where $q^0=T$, $\overline{q^{a}}$ is the mean value of $q^a$, and
$\widehat{D}_0$ stands for $\widehat{A}$.

\acknowledgments

The authors want to thank F. Barbero, J.M. Mart\'{\i}n-Garc\'{\i}a, and
E.J.S. Villase\~{n}or for helpful conversations. This work was
supported by funds provided by the Spanish MCYT projects
BFM2002-04031-C02 and BFM2001-0213.


\begin{thebibliography}{10}
\bibitem{nota0} Similarly, we can measure the
covariance (or the anti-commutator) of these two operators.
\bibitem{nota} When $\widehat{C}_T$ depends linearly on $T$, as it
is the case for the free systems that we will consider in this
work, the mean value obtained for the physical time by taking
quantum expectation values and weighting the parameter $T$ with
the probability density $\rho(T)$ is exactly $\langle\widehat{A}
\rangle\overline{T}+\langle\widehat{C}_{\overline{T}}\rangle$.
\bibitem{nota2} We can think, for instance, of a relativistic particle
with $\pi=\sqrt{H_0-\mu^2}$.
\end{thebibliography}
\end{document}